\documentclass[twocolumn,superscriptaddress,prl]{revtex4-1}
\usepackage{amssymb}
\usepackage{amsmath}
\usepackage{graphicx,float,calc}
\usepackage{color,bm}
\usepackage[colorlinks,urlcolor=blue,citecolor=blue,linkcolor=blue]{hyperref}
\usepackage{epstopdf}
\usepackage{amsfonts}
\begin{document}
\title{Non-chiral non-Bloch invariants and topological phase diagram in non-unitary quantum dynamics without chiral symmetry}

\author{Yue Zhang}
\thanks{These authors contribute equally to this work.}
\affiliation{Ministry of Education Key Laboratory for Nonequilibrium Synthesis and Modulation of Condensed Matter,Shaanxi Province Key Laboratory of Quantum Information and Quantum Optoelectronic Devices, School of Physics, Xi'an Jiaotong University, Xi'an 710049, China}

\author{Shuai Li}
\thanks{These authors contribute equally to this work.}
\affiliation{Ministry of Education Key Laboratory for Nonequilibrium Synthesis and Modulation of Condensed Matter,Shaanxi Province Key Laboratory of Quantum Information and Quantum Optoelectronic Devices, School of Physics, Xi'an Jiaotong University, Xi'an 710049, China}

\author{Yingchao Xu}
\affiliation{Ministry of Education Key Laboratory for Nonequilibrium Synthesis and Modulation of Condensed Matter,Shaanxi Province Key Laboratory of Quantum Information and Quantum Optoelectronic Devices, School of Physics, Xi'an Jiaotong University, Xi'an 710049, China}

\author{Rui Tian}
\affiliation{Ministry of Education Key Laboratory for Nonequilibrium Synthesis and Modulation of Condensed Matter,Shaanxi Province Key Laboratory of Quantum Information and Quantum Optoelectronic Devices, School of Physics, Xi'an Jiaotong University, Xi'an 710049, China}

\author{Miao Zhang}
\affiliation{Ministry of Education Key Laboratory for Nonequilibrium Synthesis and Modulation of Condensed Matter,Shaanxi Province Key Laboratory of Quantum Information and Quantum Optoelectronic Devices, School of Physics, Xi'an Jiaotong University, Xi'an 710049, China}

\author{Hongrong Li}
\affiliation{Ministry of Education Key Laboratory for Nonequilibrium Synthesis and Modulation of Condensed Matter,Shaanxi Province Key Laboratory of Quantum Information and Quantum Optoelectronic Devices, School of Physics, Xi'an Jiaotong University, Xi'an 710049, China}

\author{Hong Gao}
\affiliation{Ministry of Education Key Laboratory for Nonequilibrium Synthesis and Modulation of Condensed Matter,Shaanxi Province Key Laboratory of Quantum Information and Quantum Optoelectronic Devices, School of Physics, Xi'an Jiaotong University, Xi'an 710049, China}

\author{M. Suhail Zubairy}
\affiliation{Institute for Quantum Science and Engineering (IQSE) and Department of Physics and Astronomy, Texas A\&M University, College Station, TX 77843-4242, USA}

\author{Fuli Li}
\affiliation{Ministry of Education Key Laboratory for Nonequilibrium Synthesis and Modulation of Condensed Matter,Shaanxi Province Key Laboratory of Quantum Information and Quantum Optoelectronic Devices, School of Physics, Xi'an Jiaotong University, Xi'an 710049, China}

\author{Bo Liu}
\email{liubophy@gmail.com}
\affiliation{Ministry of Education Key Laboratory for Nonequilibrium Synthesis and Modulation of Condensed Matter,Shaanxi Province Key Laboratory of Quantum Information and Quantum Optoelectronic Devices, School of Physics, Xi'an Jiaotong University, Xi'an 710049, China}

\begin{abstract}
The non-Bloch topology leads to the emergence of various counter-intuitive phenomena in non-Hermitian systems under the open boundary condition (OBC), which can not find a counterpart in Hermitian systems. However, in the non-Hermitian system without chiral symmetry, being ubiquitous in nature, exploring its non-Bloch topology has so far eluded experimental effort. Here by introducing the concept of non-chiral non-Bloch invariants, we theoretically predict and experimentally identify the non-Bloch topological phase diagram of a one-dimensional (1D) non-Hermitian system without chiral symmetry in discrete-time non-unitary quantum walks of single photons. Interestingly, we find that such topological invariants not only can distinguish topologically distinct gapped phases, but also faithfully capture the corresponding gap closing in open-boundary spectrum at the phase boundary. Different topological regions are experimentally identified by measuring the featured discontinuities of the higher moments of the walker's displacement, which amazingly match excellently with our defined non-Bloch invariants. Our work provides a useful platform to study the interplay among topology, symmetries and the non-Hermiticity.
\end{abstract}

\maketitle

Non-Hermiticity is ubiquitous in nonconservative systems, such as in open quantum systems~\cite{Rotter2009}, correlated electron systems~\cite{Shen2018}, and systems with gain or loss~\cite{Shen2018,Konotop2016,Regensburger2012,Zhen2015,Chen2017,Zhou2018,Tang2020,Zhang2020,Ji2020}. Distinct from conventional Hermitian Hamiltonians, non-Hermitian matrices exhibit unconventional characteristics, where eigenstates in general are nonorthogonal and a complex-energy spectrum exists~\cite{Moiseyev2011}. Unprecedented phenomena with no counterparts in Hermitian systems have been unveiled, including unidirectional invisibility~\cite{Feng2013}, exceptional-point encirclement~\cite{Gao2015,Doppler2016,Xu2016} and enhanced sensitivity~\cite{Chen2017,Hodaei2017,Wiersig2014}. In particular, recently the interplay between non-Hermiticity and topology has attracted tremendous attention. Their union has resulted in a wide variety of intriguing non-Hermitian phenomena, being discovered and reinterpreted within the framework of topology ~\cite{Bergholtz2021,Kawabata2019,YutoAshida2020}. One of the particularly captivating examples is the breakdown of conventional bulk-boundary correspondence in non-Hermitian systems under OBC~\cite{Hasan2010,Qi2011}. Recent researches establish a non-Bloch band theory. Based on a generalized Brillouin zone (GBZ), non-Bloch topological invariants are introduced and the bulk-boundary correspondence can be restored~\cite{Yao2018,Yao2018a,Yokomizo2019,Yang2020}. Such quantized topological invariants are fundamentally linked to the symmetry of the system. Specifically, much attention has been paid to the systems with chiral (or equivalent) symmetry~\cite{Xiao2020,Mochizuki2020,Mochizuki2016,Xiao2017}. However, the non-Bloch topology in non-Hermitian systems without chiral symmetry has never been experimentally demonstrated yet in any system.

\begin{figure*}[tbp]
\includegraphics[width=0.85\textwidth]{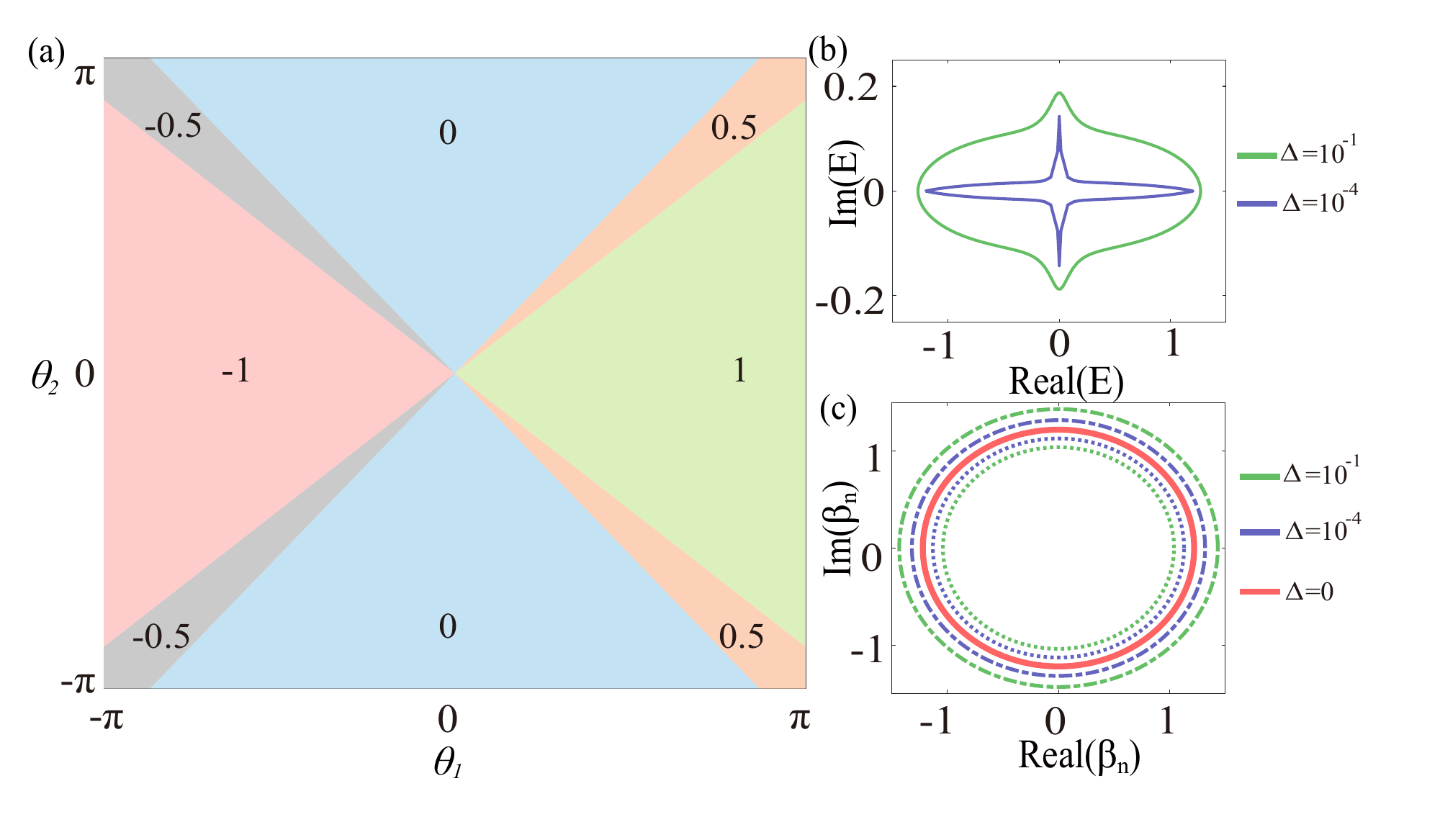}
\caption{{\bf Non-Bloch topological phase diagram of our 1D non-chiral non-Hermitian
system.} {\bf (a)} Different topological phases characterized by our defined
non-chiral non-Bloch topological invariant $v$ as the function of the coin parameter
$({\theta _1}-{\theta _2})$ with a fixed ${e^\gamma}=0.82$. The
gapped topological trivial and non-trivial regions are
characterized with $v = 0$ and $v =\pm1$, respectively. The exceptional regions are characterized with $v$ being the half integer. {\bf (b)} The bulk spectrum $E$ of our non-chiral non-unitary quantum walk in a 1D lattice under different generalized boundary conditions (GBCs) through choosing various $\Delta$. {\bf (c)} The generalized Brillouin zone (GBZ) for our non-chiral non-unitary quantum walk under different GBCs.
For each nonvanished $\Delta$, the GBZ under GBC consists of two circles. One is outside
the GBZ when $\Delta=0$ (red circle) and the other is inside that. In (b) and (c), we choose $\theta _1=0.6\pi$ and $\theta _2=0.58\pi$. Other parameters are same as in (a).}
\label{fig:fig1}
\end{figure*}

Here, we theoretically predict and experimentally identify the non-Bloch topology of a 1D non-Hermitian system without chiral symmetry in discrete-time non-unitary quantum walks of single photons. Such a study is motivated by the recent rapid progresses in exploring non-Hermitian topological phases of matter through quantum walks in various synthetic systems, including photonics~\cite{Xiao2017,Zhan2017,Wang2019,Wang2021,Xiao2021,Chen2022,Chen2023} and cold atoms~\cite{Xie2020}. Distinct from previous studies of 1D non-Hermitian system with chiral (or equivalent) symmetry~\cite{Xiao2020,Mochizuki2020,Mochizuki2016,Xiao2017},
here we explore the non-Bloch topology of non-Hermitian system without chiral symmetry. First, we introduce the concept of non-chiral non-Bloch invariants for our non-unitary quantum walks. Although the chiral symmetry is absent, quantized topological invariants can still be defined in the complex momentum plane taking into account of the deviations of localized bulk states from Bloch waves. Interestingly, we find that such topological invariants not only can characterize topologically distinct gapped phases, but also faithfully capture the corresponding gap closing  in open-boundary spectrum at the phase boundary. Therefore, the non-Bolch topological phase diagram can be unambiguously identified by them. Second, we experimentally determine the topological phase transition of our 1D non-chiral non-Hermitian system through measuring the featured discontinuities of the higher moments of the walker's displacement~\cite{Cardano2016,Cardano2017}, which amazingly match excellently with our defined non-Bloch invariants. Therefore, the non-Bolch topological phase diagram of such a non-chiral non-Hermitian system can be identified.

\textit{Non-chiral non-unitary quantum walk $\raisebox{0.01mm}{---}$} We study a non-chiral non-unitary quantum walk in a 1D lattice, which can be captured by the following Floquet operator
\begin{equation}
U_{0}=T_{\downarrow}R_{y}(\theta_{2})MT_{\uparrow}R_{y}(\theta_{1}),
\end{equation}
where ${R_y}(\theta )$  stands for the coin operator, referring to rotate coin states by   $\theta$ about the y-axis. ${T_\downarrow}({T_\uparrow})$  labels the pseudospin-dependent translation by one lattice site, where $\downarrow$  and $\uparrow$   represent the coin state $\left| 0 \right\rangle $   and  $\left| 1 \right\rangle $  , respectively. $M =\mathbb{I}_{x}\otimes(e^{\gamma}\left\vert \downarrow \right\rangle \left\langle \downarrow
\right\vert +e^{-\gamma}\left\vert \uparrow \right\rangle
\left\langle \uparrow \right\vert )$ is the polarization selective-loss operator with $\gamma$ being a tunable parameter in experiments and $\mathbb{I}_{x}$ being the identity matrix in lattice modes, which introduces the non-unitarity.

\begin{figure*}[tbp]
\includegraphics[width=0.85\textwidth]{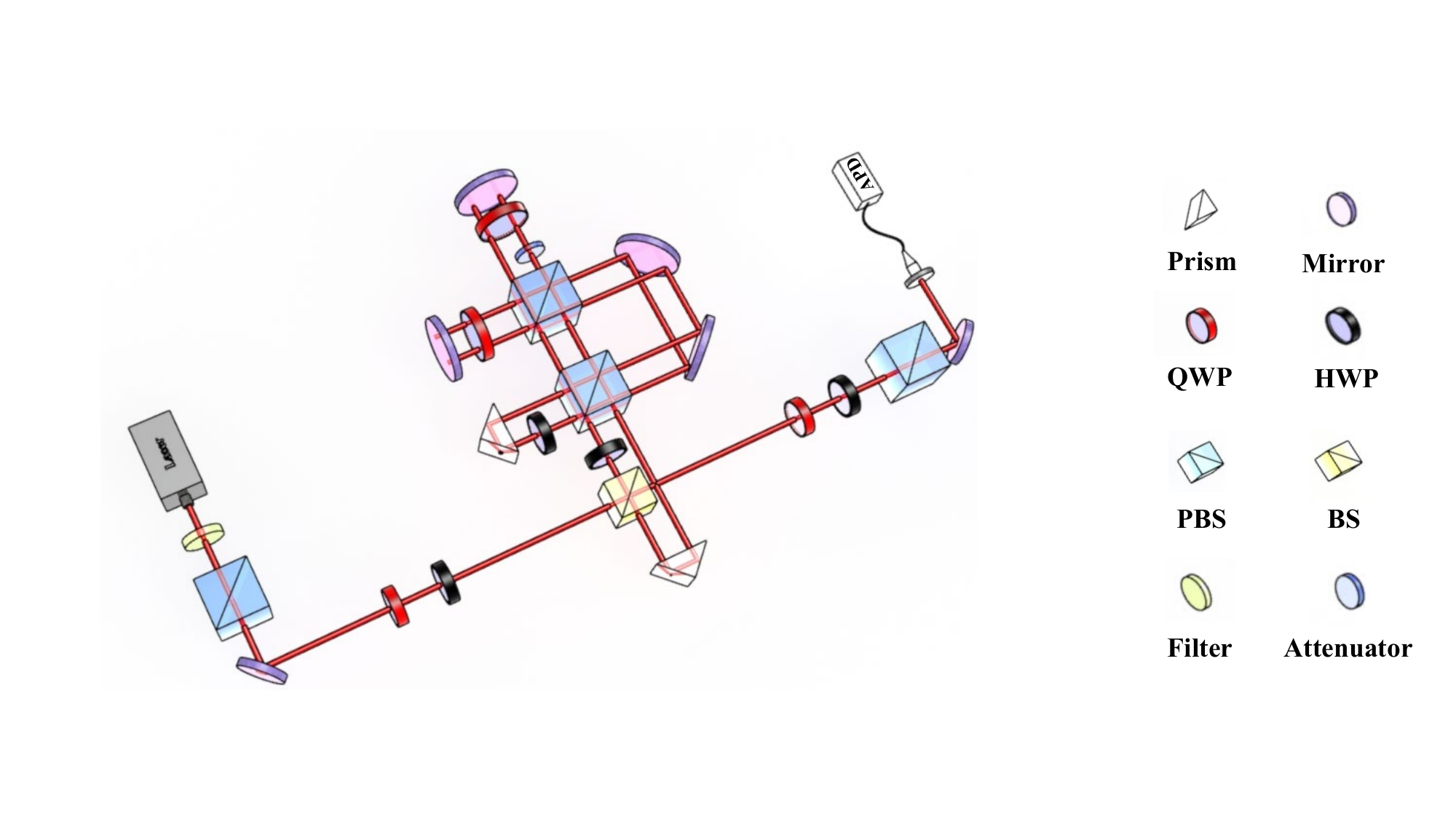}
\caption{{\bf Experimental implementation.} A laser field is attenuated to the single-photon level via neutral density filters and coupled into a polarization dependent optical loop through a beam splitter (BS) with a splitting ratio of $5/95$.
The coin operator is implemented by the half wave plate (HWP) and polarization dependent optical delay was realized by two PBS (polarizing beam splitter) loops, where the free-space path difference between horizontally and vertically polarized photons is
60 cm. M operator is implemented by adding an attenuator with attenuation rate ${e^\gamma}=0.82$ to the path of horizontally polarized photons. After using a prism to reflect the pulse back into the loop, at last around $5\%$ of photons are reflected by a beam splitter for detection and the transmitted photons continue to propagate through the interferometric network.}
\label{fig:fig3}
\end{figure*}

\begin{figure*}[tbp]
\includegraphics[width=0.8\textwidth]{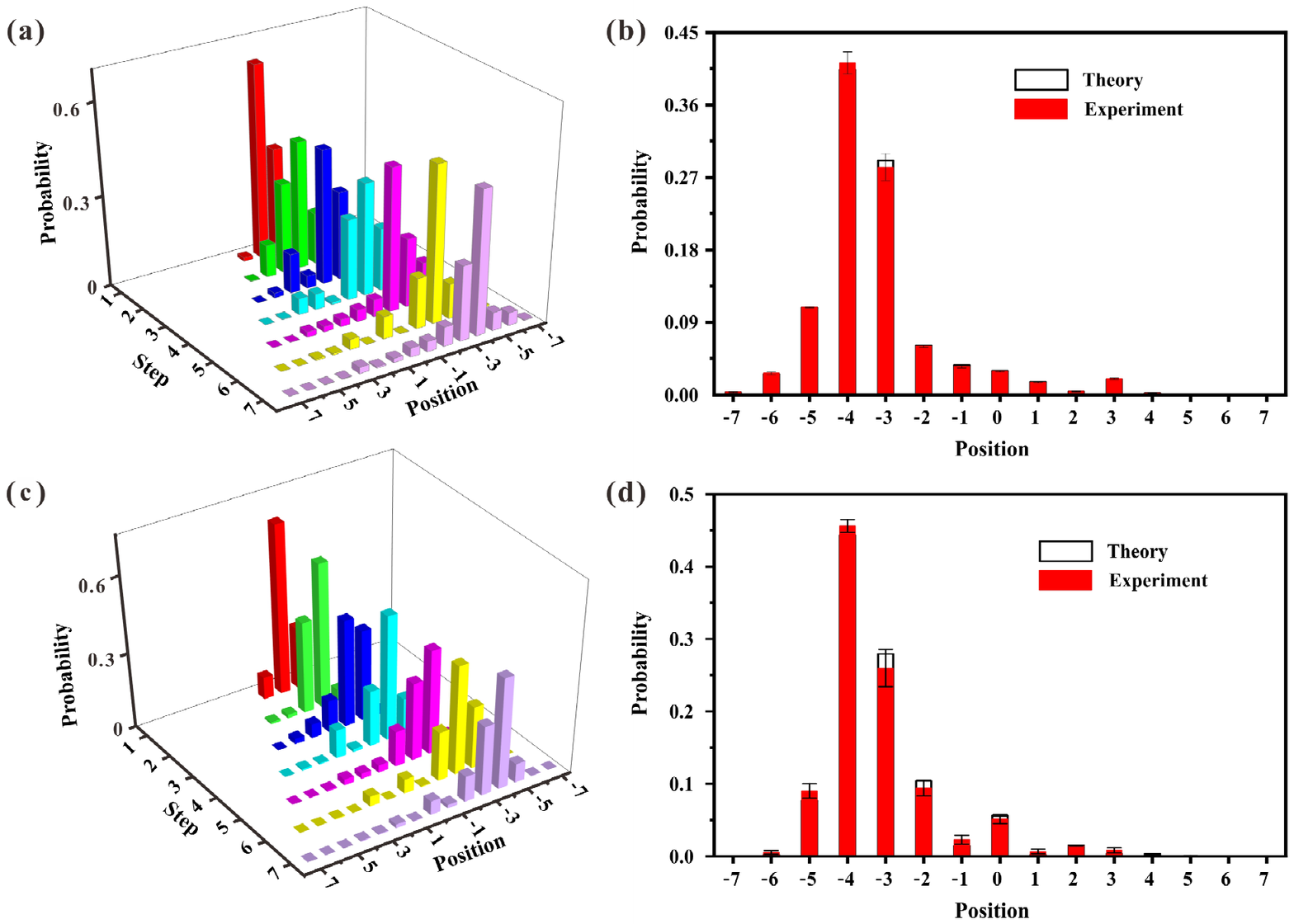}
\caption{{\bf The probability distribution for the walker.} ({\bf a}) and ({\bf c}) Time-dependent position probability distribution of a seven-step non-unitary quantum walk. ({\bf b}) and ({\bf d})
are the corresponding distribution at the last step. Here the walker is initially prepared at $x=0$ with the coin state $\left| 1 \right\rangle$. In (a) and (b), $\theta_1=0.11\pi$, while in (c) and (d), $\theta_1=0.56\pi$. Other parameters are same as in Fig. 1.}
\label{fig:fig3}
\end{figure*}

To construct the non-chiral non-Bloch band theory describing the non-unitary quantum walk governed by ${U_0}$ , we rewrite the eigenstate of ${U_0}$ as
\begin{equation}
\left\vert \psi\right\rangle =\sum\limits_{j,n=1,2}\beta_{n}^{j}\left\vert
j\right\rangle \otimes\left\vert \phi_{n}\right\rangle,
\end{equation}
where ${\beta _n}$ is the spatial-mode function for the $n$-th mode, which takes into account of the generic deviations of localized bulk states from Bloch waves. $\left| {{\phi _n}} \right\rangle $ is the corresponding coin state and $j$ stands for the lattice site. For the non-unitary quantum walk considered here, it is shown that there are two spatial modes labeled by $n = 1,2$ for each bulk state (see details in Supplementary Material(SM)). A Floquet system captured by the time-evolution operator ${U_0}$ corresponds to an effective non-Hermitian Hamiltonian ${H_{eff}}$  via the relation ${U_0} = {e^{-i{H_{eff}}}}$ , where quantum walk dynamics under ${U_0}$ can be regarded as a stroboscopic simulation of the non-unitary dynamics driven by ${H_{eff}}$. Following the non-Bloch band theory under OBC, we can express ${H_{eff}}$ on the GBZ. This can be achieved by replacing the Bloch phase factor ${e^{ik}}$ with $\beta$ on the GBZ, which manifests itself as a closed loop on the complex plane. $\beta$ can thus be regarded as the generalized quasi-momentum in the non-Bloch band theory~\cite{Yao2018,Yao2018a,Yokomizo2019,Yang2020}. Then, ${H_{eff}}$ can be expressed as  $\mathbf{\tilde{h}}(\beta)\cdot\mathbf{\sigma}$  with $\mathbf{\sigma}$ being the Pauli vector (see details in SM) and
\begin{align}
\tilde{h}_{x}^{{}}  &  =-\frac{E}{2i\sin E}(\beta a-\beta^{-1}a^{-1})\cos\theta_{2}/2 \nonumber \\
\tilde{h}_{y}  &  =-\frac{E}{2i\sin E}[i(a^{-1}+a)\cos\theta_{1}/2\sin\theta
_{2}/2 \nonumber \\
&+i(\beta a+\beta^{-1} a^{-1})\sin\theta_{1}/2\cos\theta_{2}/2] \nonumber \\
\tilde{h}_{z}  &  =\frac{E}{2i\sin E}(a-a^{-1})\sin\theta_{2}/2
\end{align}
with $ E\equiv E_{\pm} =\pm\arccos(-\frac{\alpha+\alpha^{-1}}{2}\sin\theta_{1}%
/2\sin\theta_{2}/2 +\frac{\beta\alpha+\beta^{-1}\alpha^{-1}}{2}\cos\theta_{1}/2\cos\theta
_{2}/2)$ and $a=e^{\gamma}$. Distinct from previous studies~\cite{Xiao2020,Mochizuki2020,Mochizuki2016,Xiao2017}, all the three components, i.e., ${\tilde{h}_{x(y,z)}}$ of ${\bf \tilde{h}}(\beta )$, are nonvanishing and the chiral symmetry is absent here.

\textit{ Non-chiral non-Bloch topological invariants $\raisebox{0.01mm}{---}$} To find out the topological invariant, that not only can pick out topologically distinct gapped phases, but also can determine the phase boundary through the corresponding gap closing
in open-boundary spectrum for our non-unitary quantum walks, we consider adding the generalized boundary condition (GBC) ~\cite{Guo2021}, which can be captured by the following pseudospin-dependent translation
\begin{equation*}
T_{\uparrow}^{\prime}=T_{\uparrow}+\Delta\left\vert 1\right\rangle
\left\langle L\right\vert \otimes\left\vert \uparrow\right\rangle \left\langle
\uparrow\right\vert
\end{equation*}
\begin{equation}
T_{\downarrow}^{\prime}=T_{\downarrow}+\Delta\left\vert L\right\rangle
\left\langle 1\right\vert \otimes\left\vert \downarrow\right\rangle
\left\langle \downarrow\right\vert
\end{equation}
where $\Delta $ is a real number, describing the position shift at the boundary. Then, our non-chiral non-unitary quantum walk in a 1D lattice under GBC can be captured by the following Floquet operator
\begin{equation}
U=T_{\downarrow}^{\prime}R_{y}(\theta
_{2})MT_{\uparrow}^{\prime}R_{y}(\theta_{1}).
\end{equation}
It is worthy to note that when $\Delta  = 0,U = {U_0}$ describes our studied non-chiral non-unitary quantum walk under OBC. While $\Delta  = 1$, $U$ describes the system under the periodic boundary condition (PBC). Therefore, when tuning $\Delta$ in $U$, we can continuously change the boundary condition from PBC to OBC. As shown in Fig. 1(b), the continuum spectrum of the effective Hamiltonian determined by $U$ can be obtained through diagonalising $U$ in the real space. Then, we can obtain the spatial-mode function ${\beta _n}$ associated with the corresponding continuum spectrum $E$ through the following relation (see details in SM)

\begin{equation}
\beta_{n}^{2}-\frac{2(\cos E+\frac{\alpha+\alpha^{-1}}{2}\sin\theta_{1}%
/2\sin\theta_{2}/2)}{\alpha\cos\theta_{1}/2\cos\theta_{2}/2}\beta_{n}%
+\alpha^{-2}=0
\end{equation}
Using the obtained ${\beta _n}$, the GBZ under GBC can be constructed. As shown in
Fig. 1(c), for each nonvanished $\Delta$ , the spatial-mode function  ${\beta _n}$ associated with the continuum spectrum forms two circles, construing the corresponding GBZ under GBC. When $\Delta  \to 0$, these GBZ will form two paths towards the GBZ (red circle in Fig. 1(c)) under OBC: (i) the path $C^{\text {outside}}_{\beta}$ outside
the circle of GBZ under OBC, (ii) the path $C^{\text {inside}}_{\beta}$ inside the circle of GBZ under OBC. Then, the non-chiral non-Bloch topological invariant to describe our non-unitary quantum walk captured by ${U_0}$ under OBC can be defined as

\begin{equation}
v=\frac{1}{2\pi}\sum\nolimits_{\pm}\oint_{C^{\text {inside}}_{\beta}(\Delta \rightarrow 0)}%
d\beta\left\langle \psi_{\pm}^{L}\right\vert i\partial_{\beta}\left\vert
\psi_{\pm}^{R}\right\rangle
\end{equation}
where
$
\left\vert \psi_{\pm}^{R}\right\rangle =\frac{[\sin E_{\pm}^{{}}(E_{\pm}-\sin
2E_{\pm}h_{y})+\sin E_{\pm}^{\ast}(E_{\pm}+\sin2E_{\pm}h_{y})]^{1/2}}{E_{\pm}%
\sqrt{2(E_{\pm}-h_{z})}}$
$(h_{x}-ih_{y},E_{\pm}-h_{z})^{T}%
$
and
$
\left\langle \psi_{\pm}^{L}\right\vert  =\frac{[\sin E_{\pm}^{{}}(E_{\pm}-\sin
2E_{\pm}h_{y})+\sin E_{\pm}^{\ast}(E_{\pm}+\sin2E_{\pm}h_{y})]^{-1/2}}{\sqrt
{2(E_{\pm}-h_{z})}}(h_{x}+ih_{y}$
$,E_{\pm}-h_{z})
$
are the eigenstates of the effective non-Hermitian Hamiltonian
determined by $U$. Interestingly, we find that the above defined quantized topological invariant not only can characterize topologically distinct gapped phases, but also faithfully capture the corresponding gap closing in open-boundary spectrum at the phase boundary for our non-chiral non-unitary quantum walk. As shown in Fig. 1(a), we show the non-Bloch topological phase diagram on the ${\theta _1}-{\theta _2}$  plane characterized with our defined topological invariant $v$. It is shown that the gapped topological trivial and non-trivial regions are characterized with $v=0$ and $v=\pm1$, respectively.  The gap closing in open-boundary spectrum at the phase boundary can also be faithfully captured by the transition of $v$ from integer to half integer. The exceptional regions are thus characterized with $v$ being the half integer. It is worth noting that one can also choose the outside path $C^{\text {outside}}_{\beta}$ to define the topological invariant. The same results will be obtained, i.e., the topological trivial, non-trivial and exceptional regions are characterized with $v$ being 0, $\pm1$ and half integer, respectively.

\begin{figure*}[tbp]
\includegraphics[width=0.85\textwidth]{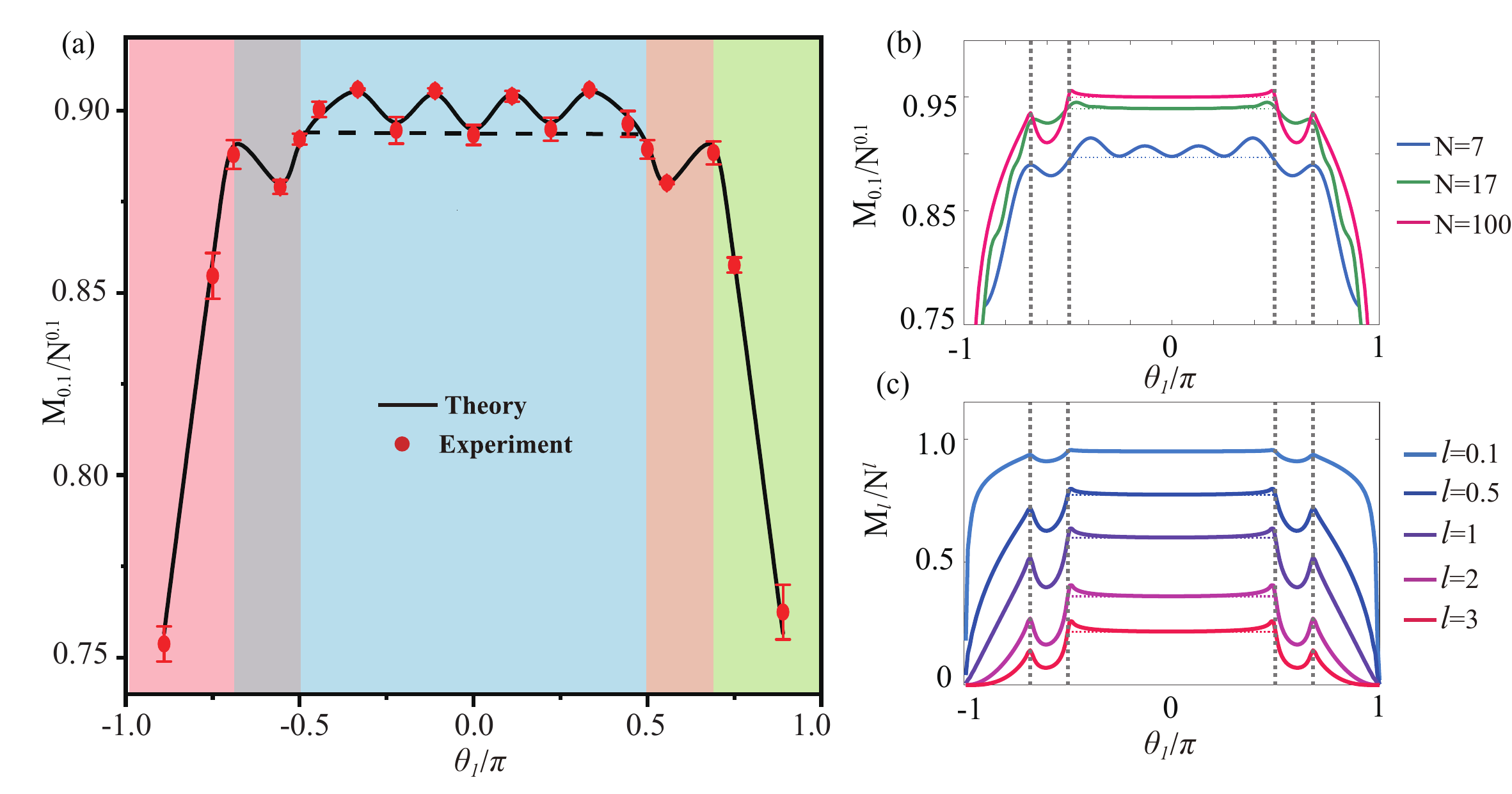}
\caption{{\bf The probability distribution moments reveal non-chiral non-Bloch topological phase transitions.} ({\bf a}) Measured probability distribution moments (red dots)
$M_{0.1}/N^{0.1}$ after a seven-step quantum walk as a function of ${\theta _1}$. Solid line represents the corresponding results of numerical simulations. The dashed line is determined by the asymptotical result as $N \to \infty$. The emergence of the abrupt slope variation
appears at each transition point between different regions
characterized with distinct non-chiral non-Bloch topological invariants $v$, which can thus be considered as an experimental visible signature of the underlying topological phase transition. ({\bf b}) Numerical simulations of $M_{0.1}/N^{0.1}$ as a function of ${\theta _1}$. It is shown that, for $M_{0.1}$, even considering small steps quantum-walk, for instance, $N=7$, the slope discontinuity of that is consistent with the asymptotic limit $N \to \infty $. ({\bf c}) Numerical simulations of $M_{l}/N^{l}$ as a function of ${\theta _1}$. Here we choose $N^{l}$ as the normalization factor and $N=80$. It is shown that different probability distribution moments $M_{l}$ with distinct $l$ show the same slope discontinuity in the asymptotic limit, which thus reveal the same topological phase transition points. Other parameters are same as in Fig. 3.}
\label{fig:fig2}
\end{figure*}

\textit{Experimental non-Bloch topological phase diagram $\raisebox{0.01mm}{---}$} We experimentally investigate the non-unitary quantum-walk dynamics governed by ${U_0}$ through implementation of a photonic discrete-time quantum walk (DTQW) via temporal-mode multiplexing scheme~\cite{Schreiber2011}. The experimental setup is illustrated in Fig. 2. The position of the walker (photon) is encoded into the arrival time at the detector. It is coupled to the coin state
realized by the horizontally and vertically polarized photons, respectively, which are labeled with $\left| 0 \right\rangle $ and $\left| 1 \right\rangle $. The $M$ operator is implemented by adding an attenuator to the path of horizontally polarized photons (see details in SI). In our experimental setup ${e^\gamma }$ in $M$ is $0.82$ .

To capture the topological quantum transition for our non-chiral non-unitary quantum walk, we employ the method of monitoring the probability distribution moments of the walker position, where a slope discontinuity demonstrates the transition point [39-40]. The moments of the probability distribution $P(m)$ associated with the walker position $m$ are defined as  $M_{l}=\sum\limits_{m}|m|^{l}P(m)$ . As shown in Fig. 4(c), the numerical simulations show that in the infinite-steps-limit, when varying $\theta_1$, at each phase transition point between different regions (in Fig. 1(a)) characterized with distinct $v$, ${M_l}$ undergoes a abrupt slope variation. Therefore, the phase boundary determined by distinct $v$ is consistent with the
slope discontinuity of ${M_l}$. For a finite number of steps, we find that ${M_l}$ have a continuous behavior converging to the asymptotical result as the step $N \to \infty $. In particular, as shown in Fig. 4(b), for $M_{0.1}$, this convergence is visible for values of $N$ that are small enough to be achieved in our non-chiral non-unitary quantum-walk.

In experiments, we study seven-step quantum-walk dynamics and focus on experimentally determining the topological phase transition of our non-chiral non-unitary quantum-walk. Initializing the walker at $x=0$ with the coin state $\left| 1 \right\rangle $, we measure the position probability distributions of a seven-step quantum walk, both the time-dependent probability distribution and the distribution at the last step are shown in Fig. 3. Based on that, we can experimentally determine ${M_{l}}$. As shown in Fig. 4(a), experimental data matched well with the numerical simulations. As expected in numerics (Fig. 4(b)), even considering seven-step quantum-walk, i.e., $N=7$, the slope discontinuity of $M_{0.1}$ is consistent with the asymptotic limit $N \to \infty $. Therefore, in our experimental detection as shown in Fig. 4(a), when varying ${\theta _1}$ with a certain ${\theta _2}$, the emergence of the abrupt slope variation in $M_{0.1}$ appears at each transition point between different regions (marked by different colors) characterized with distinct non-chiral non-Bloch topological invariants $v$, which is amazingly consistent with the phase boundary determined by $v$. The observed non-analyticity in $M_{0.1}$ can thus be considered as an experimental visible signature of the underlying topological quantum transitions. Therefore, when further varying ${\theta _2}$,  the phase diagram on the ${\theta _1}-{\theta _2}$  plane as shown in Fig. 1(a) can be experimentally identified.

\textit{Discussion} \& \textit{Conclusion} $\raisebox{0.01mm}{---}$ We have introduced the concept of non-chiral non-Bloch invariants. Through using such topological invariants, we have experimentally identified the non-Bloch topological phase diagram of a 1D non-Hermitian system without chiral symmetry in discrete-time non-unitary quantum walks of single photons. Even the chiral symmetry is absent, the quantized non-chiral non-Bloch invariants can still be defined. It not only can characterize topologically distinct gapped phases, but also faithfully capture the corresponding gap closing in open-boundary spectrum at the phase boundary. Through investigating the featured discontinuities of the higher moments of the walker's displacement, a non-chiral non-Bloch topological phase diagram for our non-unitary quantum walks is experimentally identified, amazingly matched with our defined non-Bloch invariants excellently. In prospect, our approach should be valuable for advancing the understanding of topological phenomena in open systems.

\textit{Acknowledgment $\raisebox{0.01mm}{---}$} This work is supported by the National Key R$\&$D Program of China (2021YFA1401700), NSFC (Grants No. 12074305, 12147137, 11774282), the Fundamental Research Funds for the Central Universities (Grant No. xtr052023002) and the Shaanxi Fundamental Science Research Project for Mathematics and Physics (Grant No. 23JSZ003). We also thank the HPC platform of Xi'An Jiaotong University, where our numerical calculations was performed.

\bibliographystyle{apsrev}
\bibliography{QW}

\onecolumngrid

%%%%%%%%%%%%%%%%%%%%%%%%%%%%%%%%%%%%%%
%%   Supplementary Information
%%%%%%%%%%%%%%%%%%%%%%%%%%%%%%%%%%%%%%
%\appendix
\renewcommand{\thesection}{S-\arabic{section}}
\setcounter{section}{0}  %  this will re-count section from 1
\renewcommand{\theequation}{S\arabic{equation}}
\setcounter{equation}{0}  %  this will re-count eq from 1
\renewcommand{\thefigure}{S\arabic{figure}}
\setcounter{figure}{0}  %  this will re-count eq from 1

\indent

\begin{center}\large
\textbf{Supplementary Material}
\end{center}

\subsection{Experimental implementation}

Our experimental setup is sketched in Fig. 2. Quantum walks are implemented via the temporal-mode multiplexing scheme. The photon's wave packet is provided by a pulsed laser source with central wavelength of $808$ $nm$, pulse width of $64$ $ps$, and repetition rate of $31.25$ $kHz$. The pulses are attenuated to the single-photon level by using neutral density filters, ensuring a negligible probability of multi-photon events~\cite{Schreiber2010,Schreiber2012}. The polarization-dependent translation operator is realized in the temporal domain through building the path-dependent time delays into two different paths for horizontally and vertically polarized photons, respectively. We experimentally realize a polarization
selective-loss operator $M_p =\mathbb{I}_{x}\otimes(e^{2\gamma} \left\vert \downarrow\right\rangle
\left\langle \downarrow\right\vert +
\left\vert \uparrow\right\rangle \left\langle
\uparrow\right\vert)$ through adding an attenuator to the path of horizontally polarized photons. Since $M=e^{-\gamma}M_p$, it is straightforward to map the experimentally implemented dynamics to those under $U_0$ by multiplying a time dependent factor $e^{-\gamma t}$. The coin states are encoded in the photon polarizations and the corresponding coin operator is implemented through the half-wave plates (HWPs), which can provide a careful control over the parameters $\theta_1$ and $\theta_2$. The probability distribution of the walker is measured by employing the avalanche photo-diode
(APD) to record the information regarding the number of time steps as well as the number of photons.

\subsection{ The split-step quantum walk}
Our studied non-chiral non-unitary quantum walk in a 1D lattice is implemented through the so-called split-step quantum walk scheme~\cite{Kitagawa2012,Kitagawa2010}, which can be described through the following unitary evolution operator $U_{0}=T_{\downarrow}R_{y}(\theta_{2})MT_{\uparrow}R_{y}(\theta_{1})$.  $R_{y}(\theta)\equiv \mathbb{I}_{x}\otimes e^{-i\theta\sigma_{y}/2}$ represents the rotation of the coin state about the y axis, with $\mathbb{I}_{x}\equiv\sum\limits_{j}\left\vert j\right\rangle \left\langle j\right\vert $  being the identity matrix in lattice modes. The pseudospin-dependent translation by one lattice site is captured by $T_{\uparrow}$($T_{\downarrow}$) defined as follows:

\begin{eqnarray}
\begin{split}
T_{\uparrow} & \equiv\sum\limits_{j}\left\vert j\right\rangle \left\langle
j-1\right\vert \otimes\left\vert \uparrow\right\rangle \left\langle
\uparrow\right\vert +\mathbb{I}_{x}\otimes\left\vert \downarrow\right\rangle
\left\langle \downarrow\right\vert \\
T_{\downarrow} & \equiv\sum\limits_{j}\left\vert j\right\rangle \left\langle
j+1\right\vert \otimes\left\vert \downarrow\right\rangle \left\langle
\downarrow\right\vert +\mathbb{I}_{x}\otimes\left\vert \uparrow\right\rangle
\left\langle \uparrow\right\vert
\end{split}
\end{eqnarray}
where $j$ is the site index of 1D lattice and $\downarrow$ and $\uparrow$ represent the coin state $\left\vert 0\right\rangle $ and $\left\vert 1\right\rangle $, respectively.  $M$ is the polarization selective-loss operator with $\gamma$ being a tunable parameter in experiments.

\subsection{The bulk of the non-chiral non-unitary quantum walk}

To investigate the bulk of our 1D non-chiral non-unitary quantum walk, we rewrite the bulk eigenstate $\left\vert \psi\right\rangle$ as

\begin{equation}
\left\vert \psi\right\rangle =\sum\limits_{j,n}\beta_{n}^{j}\left\vert
j\right\rangle \otimes\left\vert \phi_{n}\right\rangle
\end{equation}
where ${\beta _n}$ is the spatial-mode function for the $n$-th mode, which takes into account the generic deviations of localized bulk states from Bloch waves. ${\phi _n}$ is the corresponding coin state and $j$ is the site index of 1D lattice. The bulk
of the 1D chain can be investigated through solving the following
eigenstate equation

\begin{equation}
(A_{m}\beta_{n}+A_{p}\beta_{n}^{-1}+A_{s}-\lambda)\left\vert \phi_{n}\right\rangle
=0
\end{equation}
where $A_{m}=F_{m}MG_{s}$ , $A_{p}=F_{s}MG_{p}$ and $A_{s}   =F_{s}MG_{s}+F_{m}MG_{p}$ with $F_{m}    =P_{\downarrow}R_{y}(\theta_{2})$ , $F_{s}    =P_{\uparrow}R_{y}(\theta_{2})$ , $G_{s}    =P_{_{\downarrow}}R_{y}(\theta_{1})$ and  $G_{p}   =P_{\uparrow}R_{y}(\theta_{1})$. Here   $P_{\uparrow}=\left\vert \uparrow\right\rangle \left\langle \uparrow
\right\vert $, $P_{_{\downarrow}}=\left\vert \downarrow\right\rangle
\left\langle \downarrow\right\vert $ and $\lambda=\exp(-iE)$. To make the above equation have non-trivial solutions, the following relation should be satisfied
\begin{equation}
\det(A_{m}\beta_{n}+A_{p}\beta_{n}^{-1}+A_{s}-\lambda)=0
\end{equation}
Through solving Eq. (S3), we find that there exist two solutions of ${\beta _n}$ with $n=1,2$ and Eq. (S3) can thus be rewritten as

\begin{equation*}
(A_{m}\beta_{1}+A_{p}\beta_{1}^{-1}+A_{s}-\lambda)\left\vert \phi
_{1}\right\rangle =0
\end{equation*}

\begin{equation}
(A_{m}\beta_{2}+A_{p}\beta_{2}^{-1}+A_{s}-\lambda)\left\vert \phi
_{2}\right\rangle =0
\end{equation}
The energy spectrum can thus be obtained as
\begin{eqnarray}
\begin{split}
E \equiv E_{\pm} &  =\pm\arccos(-\frac{\alpha+\alpha^{-1}}{2}\sin\theta_{1}%
/2\sin\theta_{2}/2+\frac{\beta_{n}\alpha+\beta_{n}^{-1}\alpha^{-1}}{2}\cos\theta_{1}/2\cos\theta
_{2}/2)
\end{split}
\end{eqnarray}
The effective non-Hermitian Hamiltonian constructed from the above equation Eq. (S4) can be expressed in the form  $\mathbf{h\cdot\sigma}$

\begin{eqnarray}
\begin{split}
h_{x}^{{}}  &  =-\frac{E}{2i\sin E}(\beta_{n} a-\beta_{n}^{-1}a^{-1})\cos\theta_{2}/2\\
h_{y}  &  =-\frac{E}{2i\sin E}[i(a^{-1}+a)\cos\theta_{1}/2\sin\theta
_{2}/2+i(\beta_{n} a+\beta_{n}^{-1} a^{-1})\sin\theta_{1}/2\cos\theta_{2}/2]\\
h_{z}  &  =\frac{E}{2i\sin E}(a-a^{-1})\sin\theta_{2}/2
\end{split}
\end{eqnarray}
The corresponding eigenstates can be obtained
\begin{equation}
\left\vert \psi_{\pm}^{R}\right\rangle =\frac{[\sin E_{\pm}^{{}}(E_{\pm}-\sin
2E_{\pm}h_{y})+\sin E_{\pm}^{\ast}(E_{\pm}+\sin2E_{\pm}h_{y})]^{1/2}}{E_{\pm}
\sqrt{2(E_{\pm}-h_{z})}}(h_{x}-ih_{y},E_{\pm}-h_{z})^{T}%
\end{equation}
\begin{equation}
\left\langle \psi_{\pm}^{L}\right\vert =\frac{[\sin E_{\pm}^{{}}(E_{\pm}-\sin
2E_{\pm}h_{y})+\sin E_{\pm}^{\ast}(E_{\pm}+\sin2E_{\pm}h_{y})]^{-1/2}}{\sqrt
{2(E_{\pm}-h_{z})}}(h_{x}+ih_{y},E_{\pm}-h_{z})
\end{equation}

Since the OBC imposes one further condition $\left| {{\beta _1}} \right| = \left| {{\beta _2}}\right|$ for the bulk modes, these spatial modes can be labeled by $\beta$, which will form the GBZ as shown in the red circle in Fig. 1(c). Through Eq. (S6), we can obtain the effective non-Hermitian Hamiltonian ${H_{eff}}$ as shown in Eq. (3) in the main text.

\subsection{Constructing GBZ under the generalized boundary condition (GBC)}

To construct our proposed non-chiral non-Bloch topological invariants, we consider imposing the generalized boundary condition (GBC) to our system, which can be captured by the following Floquet  operator   $U=T_{\downarrow}^{\prime}R_{y}(\theta
_{2})MT_{\uparrow}^{\prime}R_{y}(\theta_{1})$. To find out the generalized Brillouin zone (GBZ) under GBC, i.e., considering $U$ with a certain novanished $\Delta $ , we diagonalize $U$ in its real space matrix form
$\left(
\begin{array}
[c]{cc}%
U_{i,j}^{11} & U_{i,j}^{12}\\
U_{i,j}^{21} & U_{i,j}^{22}%
\end{array}
\right)$ with $U_{i,j}^{11}=(\delta_{i+1,j}+\delta_{i,L}\delta_{1,j}\Delta)e^{\gamma}%
\cos\theta_{1}/2\cos\theta_{2}/2-(\delta_{i,j}+\delta_{i,L}\delta_{L,j}%
\Delta^{2})e^{-\gamma}\sin\theta_{1}/2\sin\theta_{2}/2$, $U_{i,j}^{22}=(\delta_{i,j+1}+\delta_{i,1}\delta_{L,j}\Delta)e^{-\gamma}%
\cos\theta_{1}/2\cos\theta_{2}/2-\delta_{i,j}e^{\gamma}\sin\theta_{1}%
/2\sin\theta_{2}/2$, $U_{i,j}^{21}=(\delta_{i,j+1}+\delta_{i,1}\delta_{L,j}\Delta)e^{-\gamma}%
\sin\theta_{1}/2\cos\theta_{2}/2+\delta_{i,j}e^{\gamma}\cos\theta_{1}%
/2\sin\theta_{2}/2$ and $U_{i,j}^{12}=-(\delta_{i+1,j}+\delta_{i,L}\delta_{1,j}\Delta)e^{\gamma}%
\sin\theta_{1}/2\cos\theta_{2}/2-(\delta_{i,j}+\delta_{i,L}\delta_{L,j}%
\Delta^{2})e^{-\gamma}\cos\theta_{1}/2\sin\theta_{2}/2$. The continuum spectrum can thus be found out. Through substituting the quasienergy of the continuum spectrum into Eq. (S5), the corresponding spatial-mode function  ${\beta _n}$ can be obtained. We thus plot these ${\beta _n}$ in the complex plane, as shown in Fig. 1(c) in the main text, the GBZ under GBC can be constructed.

\end{document}